\documentclass[twocolumn, tighten, twocolappendix]{aastex63}
\accepted{January 27, 2021}
\submitjournal{ApJ}
\shorttitle{MAPPIES II - Application to the UV/soft X-ray excess in Blazar Spectra}
\shortauthors{Dreyer and B{\"o}ttcher}
\graphicspath{{./}{figures/}}
\usepackage{amsmath}
\begin{document}

\title{Monte-Carlo Applications for Partially Polarized Inverse External-Compton Scattering (MAPPIES) II - Application to the UV/Soft X-ray Excess in Blazar Spectra}
\correspondingauthor{Lent{\'e} Dreyer}
\email{lentedreyer@gmail.com}

\author[0000-0002-4971-3672]{Lent{\'e} Dreyer}
\affiliation{Centre for Space Research, North-West University, Potchefstroom 2531, South Africa}
\author[0000-0002-8434-5692]{Markus B{\"o}ttcher}
\affiliation{Centre for Space Research, North-West University, Potchefstroom 2531, South Africa}

\begin{abstract}
    The spectral energy distributions (SEDs) of some blazars exhibit an ultraviolet (UV) and/or soft X-ray excess, which can be modelled with different radiation mechanisms. Polarization measurements of the UV/X-ray emission from blazars may provide new and unique information about the astrophysical environment of blazar jets, and could thus help to distinguish between different emission scenarios. In this paper, a new Monte-Carlo code -- MAPPIES (Monte-Carlo Applications for Partially Polarized Inverse External-Compton Scattering) -- for polarization-dependent Compton scattering is used to simulate the polarization signatures in a model where the UV/soft X-ray excess arises from the bulk Compton process. Predictions of the expected polarization signatures of Compton emission from the soft X-ray excess in the SED of AO 0235+164, and the UV excess in the SED of 3C 279 are made for upcoming and proposed polarimetry missions. 
\end{abstract}
\section{Introduction} \label{sec:INTRO}
Active galactic nuclei (AGNs) are some of the most luminous objects in the universe. About 10\% of AGNs are observed to host relativistic jets, which are considered to be powerful emitters of radiation across the entire electromagnetic spectrum. Blazars are an extreme class of AGNs -- consisting of BL Lac objects and flat-spectrum radio quasars (FSRQs) -- for which the observer's line of sight is closely aligned to the jet axis \citep{Urry_Padovani_1995, Padovani_etal2018}. Various properties of the radiation from blazars have been studied with multi-wavelength observations and spectral fitting. While FSRQs have strong optical emission lines (which indicates the presence of accretion-disk radiation and a dense broad line region (BLR)), BL Lac objects typically do not have a luminous accretion-disk or broad lines \citep{Giommi_etal2012, Dermer_Giebels_2016}. The spectral energy distributions (SEDs) of blazars are dominated by non-thermal emission, and generally consist of two distinct components; the low-frequency (radio through ultraviolet (UV) or X-ray) component, and the high-frequency (X-ray and $\gamma$-ray) component. The relativistic jets contain ultra-relativistic electrons that produce soft photons from radio frequencies up to UV/X-rays through synchrotron emission, and photons up to very high energies (VHE; $E\geq 100~\mathrm{GeV}$) via inverse-Compton (IC) processes (i.e. leptonic models). Alternatively, high-energy emission can be produced by synchrotron radiation of pair cascades, powered by hadronic processes and synchrotron emission of ultra-high-energy protons and muons (i.e. hadronic models). Both leptonic models and hadronic models are generally able to produce acceptable fits to blazar SEDs \citep{Bottcher_etal2013}. 

Polarization is an important observable that can be used to constrain the morphology and geometry of the emitting region, and to distinguish between various emission mechanisms. Polarization measurements of the radio and optical emission from blazar jets have been the key to understanding many diverse aspects of blazar jets (see e.g. \cite{Bottcher_2019, Trippe_2019, Zhang_2019} for recent reviews). The radio and optical emission of blazars have moderate polarization degrees (PDs) up to $(3-30) \%$ \citep{Conway_etal1993, Zhang_etal2014}, which correspond to non-thermal electron synchrotron emission \citep{Westfold_1959, RybickiandLightman_1979}, thus confirming the dominant radiation mechanism for the radio and optical emission from blazars. 

The polarization of the UV, X-ray and $\gamma$-ray emission has so far been largely unexplored, although its scientific potential has long been appreciated (see e.g. \cite{Krawczynski_etal2011, Andersson_etal2015, Zhang_2017, Mignani_etal2019, Rani_etal2019} for reviews). For instance, the synchrotron origin of the X-ray emission from high-synchrotron-peaked (HSP) blazars may be confirmed with X-ray polarimetry (e.g. \cite{Krawczynski_2012}). High-energy polarimetry may also be able to distinguish between leptonic and hadronic emission scenarios for the origin of the high-frequency component in blazar SEDs, since hadronic models typically predict higher degrees of X-ray and $\gamma$-ray polarization than leptonic models \citep{Zhang_Bottcher_2013, Paliya_etal2018, Zhang_etal2019}. In leptonic models, the high-energy emission can be partially polarized, depending on the source of the seed photon field for Compton scattering. While external-Compton emission is expected to be unpolarized, Compton scattering of the polarized low-frequency synchrotron emission (i.e. synchrotron-self-Compton; SSC) is expected to be polarized with a polarization degree (PD) that is about half of the polarization of the synchrotron emission (e.g. \cite{Chakraborty_etal2015}). 

In addition to the two characteristic-broad, non-thermal components described above, blazar SEDs sometimes exhibit an infrared (IR) bump, optical/UV bump (called the \textit{Big Blue Bump} (BBB)), and/or soft X-ray excess (e.g. \cite{Masnou_etal1992, Grandi_etal1997, Haardt_etal1998, Pian_etal1999, Raiteri_etal2005, Palma_etal2011, Ackermann_etal2012, Paliya_etal2015, Pal_etal2020}; see \cite{Antonucci_2002, Perlman_etal2008} for reviews). Various radiation mechanisms have been proposed for the origin of the UV/soft X-ray excess in the SEDs, which include the following:
\begin{itemize}
     \item Thermal emission from the accretion-disk (e.g. \cite{Pian_etal1999, Blaes_etal2001, Paliya_etal2015, Pal_etal2020}). \\
     \item A higher than galactic dust-to-gas ratio to the source, resulting in an over-estimation of the neutral-hydrogen column density and, therefore, an over-correction for the corresponding photo-electric absorption at low X-ray energies (in this case, the excess would not actually be physical; e.g. \cite{Ravasio_etal2003}).
     \item A distinct synchrotron component from a different region than the low-frequency component in a multi-zone construction (e.g. \cite{Paltani_etal1998, Ostorero_etal2004, Raiteri_etal2005}). 
     \item Synchrotron emission from VHE $\gamma$-ray induced pair cascades in blazar environments \citep{Roustazadeh_Botther_2012}.
     \item Reduced radiative cooling of the highest-energy electrons in a Compton-dominated blazar, due to the Klein-Nishina suppression of the Compton cross section (e.g. \cite{Ravasio_etal2003, Moderski_etal2005}).
     \item A signature of IC scattering of external radiation fields by a thermal, non-relativistic population of electrons (i.e. the bulk Compton effect; e.g. \cite{Sikora_etal1994, Sikora_etal1997, Blazejowski_etal2000, Ackermann_etal2012, Baring_etal2017}). 
 \end{itemize}
 Polarization measurements of the UV and soft X-ray emission may yield new and unique information about these spectral features in the SEDs, thus distinguishing between different emission scenarios. Given these promising prospects, numerous polarimetry missions anticipating to deliver polarization measurements of the UV/X-ray emission from astrophysical sources (including blazars) are at various stages of planning, design, and operation. It is, therefore, important to consider predictions of the polarization signatures for different emission scenarios that may be able to explain the origin of the UV/soft X-ray excess in blazar spectra. 

In this paper, a new Monte-Carlo code -- MAPPIES (Monte-Carlo Applications for Partially Polarized Inverse External-Compton Scattering; \cite{PaperI}) -- is used to simulate the polarization signatures in a model where the UV/soft X-ray excess arises due to Compton scattering of external fields by thermal electrons contained in the blazar jet, as proposed by \cite{Baring_etal2017} for the BL Lac object AO 0235+164. Two blazar case studies are presented: The BL Lac object AO 0235+164, and the FSRQ 3C 279. An overview of the soft X-ray excess in the SED of AO 0235+164 and the BBB in the SED of 3C 279 is given in Section \ref{sec:0235+164} and Section \ref{sec:3C279}, respectively. The model setup and parameters considered for the simulations are described in Section \ref{sec:parameters}, followed by the results in Section \ref{sec:results}, and the conclusions in Section \ref{sec:summary}. 
\section{The soft X-ray excess in the SED of ~ AO 0235+164}\label{sec:0235+164}
The BL Lac object AO 0235+164 (redshift $z = 0.94$) is one of the most prominent examples of the emergence of a soft X-ray excess (e.g. \cite{Junkkarinen_2004, Raiteri_etal2005, Ackermann_etal2012}). The optical to X-ray SED was reconstructed by \cite{Raiteri_etal2006} with optical through UV data from the X-ray Multi-Mirror Mission (XMM-Newton; \cite{Talavera_2009}) during 2000 - 2005. The source was mostly faint, with a hard X-ray spectrum during that time, except in February 2002, when the source was found in a bright state with a steep X-ray spectrum. The XMM-Netwon data indicated the existence of UV to soft X-ray bump in the SED with an inferred peak frequency of $\mathrm{log_{10}(\nu/ Hz)}\sim 15.5-16.1$. While thermal emission from the disk might be able to explain the bump, \cite{Raiteri_etal2006} proposed that synchrotron emission from a region that is closer to the black hole than where the low-frequency component originates, could be another possible mechanism. Additionally, \cite{Ostorero_etal2004} showed that the soft X-ray excess can be obtained in the ambit of the rotating helical jet model (e.g. \cite{Villata_Raiteri_1999}) by admitting a synchrotron contribution to the X-ray radiation. 
A soft X-ray excess within the frequency range $\mathrm{log_{10}(\nu/ Hz)}\sim 16.5-18.3$ of the SED, was reported by \cite{Ackermann_etal2012} during a multi-wavelength campaign between 2008 and 2009, when AO 0235+164 was in a flaring state. The high-frequency component of the SED was interpreted with the standard leptonic scenario (i.e. IC scattering of the external IR radiation from the dusty torus). The X-ray data from the \textit{Swift} X-ray Telescope (\textit{Swift}-XRT; \cite{Burrows_etal2005}) and the Rossi X-ray Timing Explorer (RXTE; \cite{Rothschild_etal1998}) presented a soft X-ray excess in the SED. Since the X-ray spectrum was too soft to be attributed to the SSC component, IC scattering off cold electrons in the jet was considered to be a possibility.

\begin{figure}[ht!]
\includegraphics[width=\columnwidth]{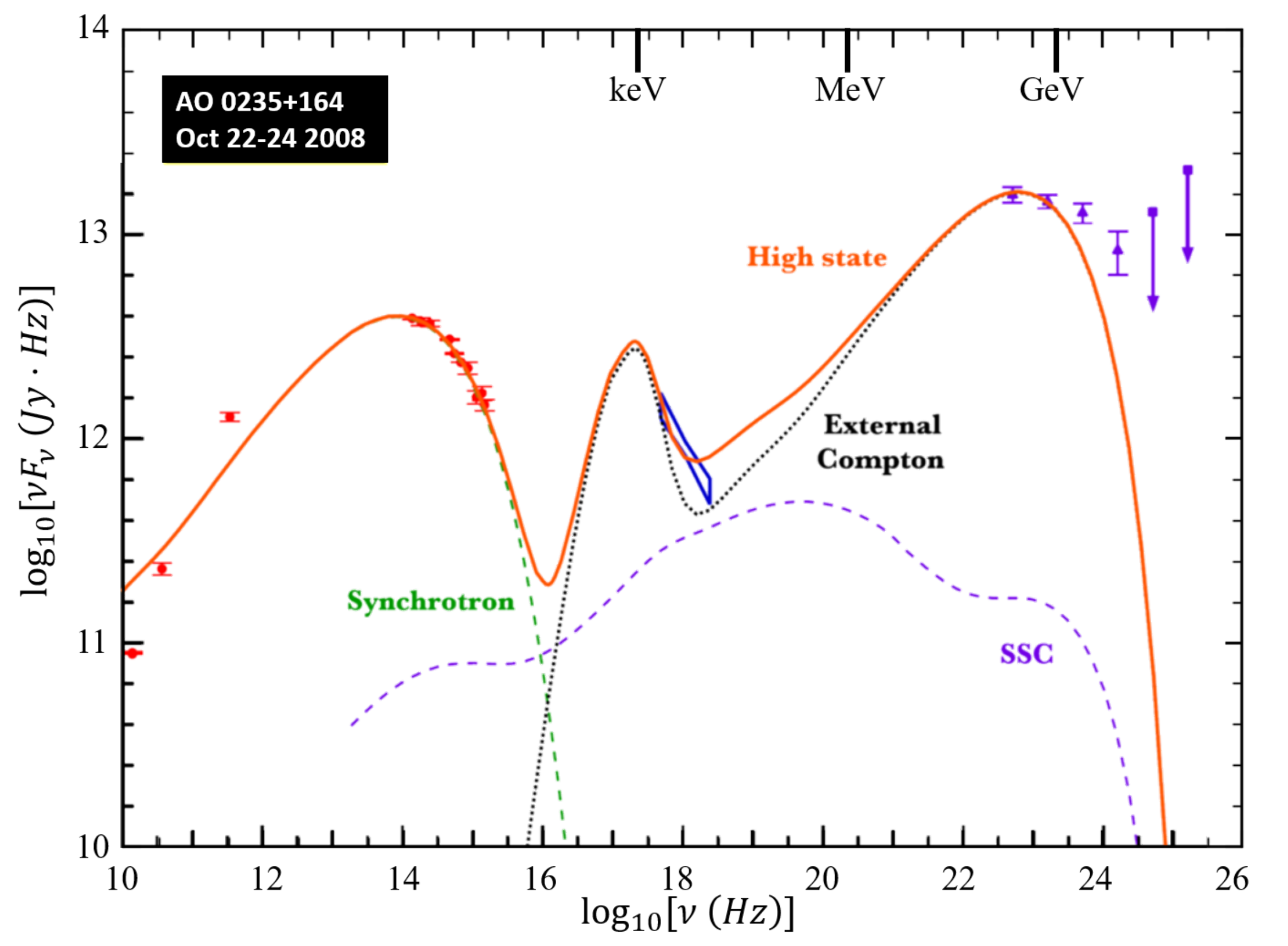}
\caption{The multi-wavelength spectrum (points) spanning the radio, optical, X-ray, and $\gamma$-ray bands, together with model fits from \cite{Baring_etal2017}, for the 2008 October high-state Fermi-LAT observation of AO 0235+164 (data from \cite{Ackermann_etal2012}). The blue \textit{butterfly} block represents Swift-XRT data. The broad-band model consists of a synchrotron component (dashed green curve) up to the optical band, a two order SSC contribution in the optical, X-rays and $\gamma$-rays (dashed orange curve), and external Compton (EC) emission, including a bulk Compton feature (dotted black curve) between $\mathrm{log_{10}(\nu/ Hz)}\sim 17.4$ and $\mathrm{log_{10}(\nu/ Hz)}\sim 24.4$. The orange curve is the total model spectrum (which includes a very small correction for $\gamma \gamma$ absorption by the extra-galactic background light). Taken from \cite{Baring_etal2017}. \label{fig:Baring_etal2017}}
\end{figure}
\cite{Baring_etal2017} employed the multi-wavelength observations of \cite{Ackermann_etal2012} and modelled the soft X-ray excess as a bulk Compton component that results from an external radiation field scattering off a thermal population of shock-heated electrons contained in the blazar jet (see Figure \ref{fig:Baring_etal2017}). Monte-Carlo simulations of the diffusive shock acceleration (DSA) process by \cite{Summerlin_Baring_2012} were coupled with the radiation transfer modules of \cite{Bottcher_etal2013}, and the fit of the soft X-ray feature through the bulk Compton process aided in fixing the thermal-to-non-thermal particle ratio in the jet, thus tightly constraining the particle diffusion parameters in the DSA process. A qualitative prediction for this scenario is that the thermal Comptonization process should lead to significant polarization in the soft X-ray spectral component. In what follows, the model of \cite{Baring_etal2017} will be considered in order to test this prediction for the soft X-ray excess in the SED of AO 0235+164, using the MAPPIES code.
\section{The BBB in the SED of 3C 279}\label{sec:3C279}
The FSRQ 3C 279 (redshift $z = 0.536$) is one of the brightest $\gamma$-ray sources in the sky, and the first blazar to be detected by the Energetic Gamma Ray Experiment Telescope (EGRET; \cite{Kanbach_etal1988, Hartman_etal1992}). Multi-wavelength variability and radio/optical polarimetry suggest that the broadband spectrum of 3C 279 at radio to UV frequencies is produced by synchrotron radiation (e.g. \cite{Maraschi_etal1994, Hartman_1996}). The UV-optical continuum usually has a steep power-law spectrum, due to radiation losses of the relativistic electrons in the jet, revealing its non-thermal origin. The source was monitored by \cite{Pian_etal1999} with the International Ultraviolet Explorer (IUE; \cite{Nichols_Linsky_1996}) -- combining the UV data with observations from the ROentgen SATellite (ROSAT;\cite{Truemper_1993}) and EGRET -- during its low state, thus allowing the detection of an excess in the UV regime at $\mathrm{log_{10}(\nu/ Hz)}\sim 15.2$ (i.e, the BBB; otherwise hidden below the dominant power-law continuum attributed to non-thermal emission from the jet). The BBB was suggested to be due to thermal emission from the accretion-disk (see also e.g. \cite{Blaes_etal2001, Pal_etal2020}), and the UV luminosity they found for the presumed accretion-disk responsible for the BBB feature was consistent with estimates provided by \cite{Dermer_Schlickeiser_1993} and \cite{Sikora_etal1994}.

\begin{figure}[ht!]
\includegraphics[width=\columnwidth]{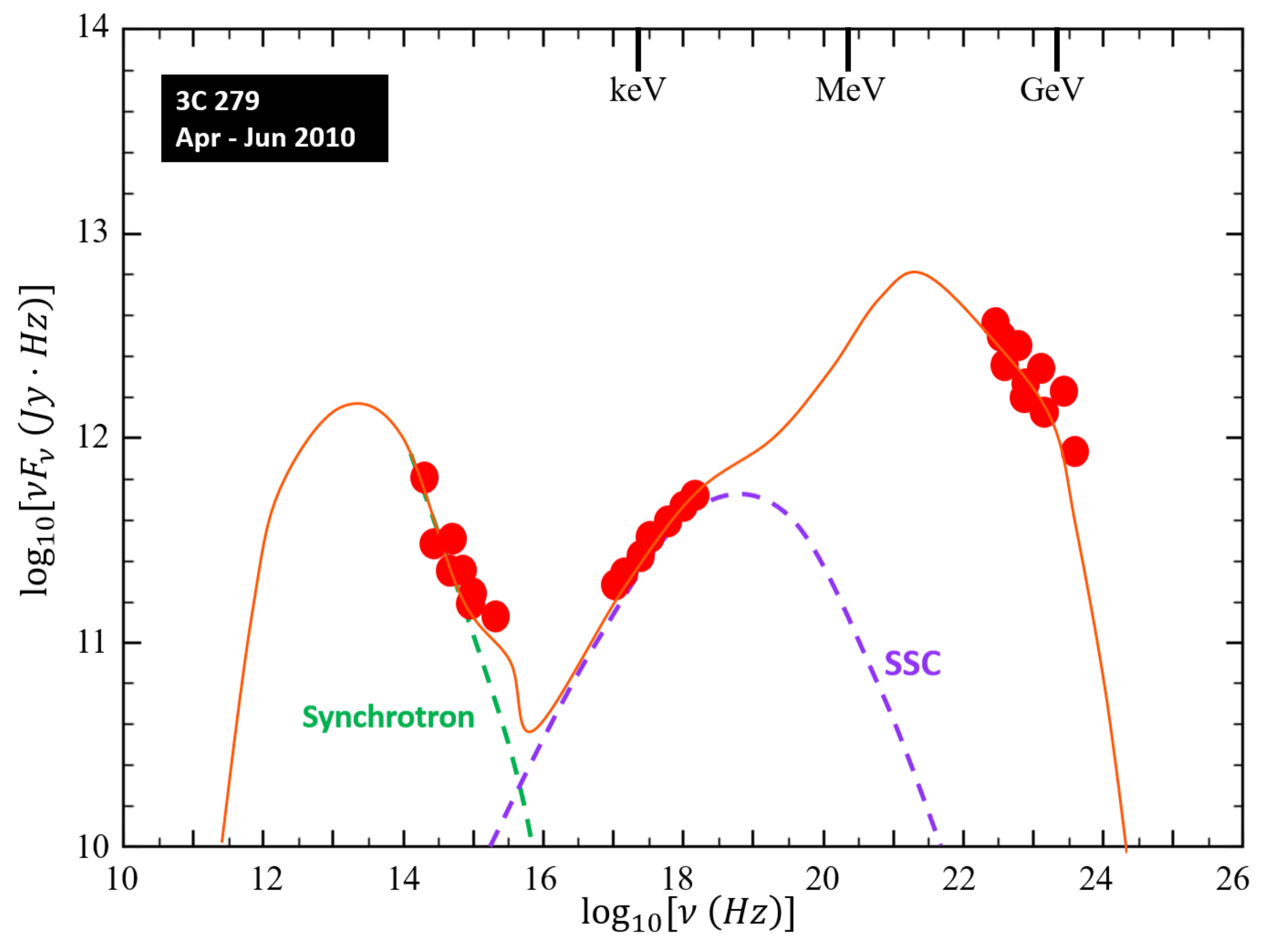}
\caption{The SED of 3C 279 during low-activity states. Simultaneous data from the Small and Moderate Aperture Research Telescope System (SMARTS), Swift XRT, and Fermi-LAT are shown with red circles. The green and orange dashed curves correspond to synchrotron and SSC emission, respectively. The orange solid curve is the sum of all the radiative mechanisms (the thermal contributions from the torus, accretion-disk, and X-ray corona, as well as the external-Compton disk, BLR, and the dusty torus components are not shown in the figure). From \cite{Paliya_etal2015}. \label{fig:Paliya_etal2015}}
\end{figure}
A detailed study of the brightest $\gamma$-ray flare observed from 3C 279 was presented by \cite{Paliya_etal2015}, and the modeling of the low-activity state showed a slight \textit{turnover} at $\mathrm{log_{10}(\nu/ Hz)}\sim 14.5 - 15.7$ (see Figure \ref{fig:Paliya_etal2015}), which was also attributed to the accretion-disk radiation. The dilution of the synchrotron polarization, due to unpolarized accretion-disk emission towards the UV, will result in a decrease of the PD throughout the optical - UV regime. Additionally, the radiation emitted by electron-scattering–dominated accretion-disks is expected to be considerably polarized, perpendicular to the disk axis, with a strong angle dependence (up to a PD $\sim 11.7\%$ for an \textit{edge-on} disk; e.g. \cite{Chandrasekhar_1960}). However, assuming that the jet of a blazar propagates along the symmetry axis of the accretion disk, the high-energy emission region in the jet will have a perfect \textit{face-on} view of the disk, in which case there is no net polarization in the accretion-disk emission due to the azimuthal symmetry \citep{Smith_etal2004}.

Synchrotron emission from VHE $\gamma$-ray induced pair cascades in blazar environments, was suggested by \cite{Roustazadeh_Botther_2012} as an alternative contribution to the BBB. This cascade emission may peak in the UV/soft X-ray range for sufficiently strong magnetic fields, and can resemble the BBB in the UV regime. The external radiation field was modelled as isotropic black body radiation with a temperature of $kT_{\mathrm{rad}} \sim 1.7$ eV. The results illustrated that synchrotron emission from VHE $\gamma$-ray induced pair cascades can reproduce the BBB in 3C 279, peaking at $\mathrm{log_{10}(\nu/ Hz)}\sim 14.9-15.3$. Within this scenario, the BBB would result in polarized emission, with a weak dependence on the frequency over the optical/UV range.
\newpage
\section{Model setup}\label{sec:parameters}
The MAPPIES code is a newly developed Monte-Carlo code for polarization-dependent Compton scattering of external fields in jet dominated astrophysical sources \citep{PaperI}. An external radiation field (originating in the AGN rest frame) scatters off an arbitrary (thermal and non-thermal) electron distribution, assumed to be isotropic in the co-moving frame of the emission region that moves along the jet with a bulk Lorentz factor $\Gamma_{\mathrm{jet}}$. The polarization signatures are calculated using the Stokes formalism \citep{Stokes_1851}, and the polarization-dependent Compton scattering of the seed photons are calculated following Monte-Carlo methods by \cite{Matt_etal1996}. The code is used to simulate the polarization signatures in a model where the UV/Soft X-ray excess in the SEDs of blazars is due to bulk Compton emission. In this section, we describe the model setup for the two case studies of AO 0235+164 and 3C 279. In particular, a description of the seed photon fields and electron energy distributions is given. 
\begin{deluxetable}{lc}
\tablenum{1}
\tablecaption{The model parameters for the BL Lac object AO 0235+164. \label{tab:param_AO235_164}}
\tablewidth{0pt}
\tablehead{\colhead{\textbf{Parameter description}} & \colhead{Value}}
\startdata
\hline
Redshift of the source, $z$ & $0.94$\\
Bulk Lorentz factor, $\Gamma_{\mathrm{jet}}$ & $35.0$\\
Luminosity distance, $d_L$ & $1.89 \times 10^{28}~\mathrm{cm}$\\
Temperature of the dusty torus, $kT_{\mathrm{rad}}$ & $0.1~\mathrm{eV}$\\
Radiation energy density of emission & \\
from the dusty torus, $U_{\mathrm{DT}}$ & $0.6~\mathrm{erg \cdot cm^{-3}}$\\
 Effective size of the dusty torus $R_{\mathrm{eff}}$\tablenotemark{a} & $6\times10^{17}~\mathrm{cm}$\\
\enddata
\tablenotetext{a}{See main text for the definition of the effective size}
\tablecomments{The model parameters correspond to those of \cite{Baring_etal2017}. The seed photon field is assumed to be IR emission from the dusty torus, and the electron energy distribution is drawn from the electron fit spectrum of \cite{Baring_etal2017} in order to simulated the IC scattering off thermal shock-heated electrons in the jet of the blazar.}
\end{deluxetable}
\begin{deluxetable}{lc}
\tablenum{2}
\tablecaption{The model parameters for the FSRQ 3C 279. \label{tab:param_3C279}}
\tablewidth{0pt}
\tablehead{\colhead{\textbf{Parameter description}} & \colhead{Value}}
\startdata
\hline
Redshift of the source, $z$ & $0.536$\\
Bulk Lorentz factor, $\Gamma_{\mathrm{jet}}$ & $10.0$\\
Luminosity distance, $d_L$ & $9.3 \times 10^{27}~\mathrm{cm}$\\
Black hole mass, $M_{\mathrm{BH}}$ & $3 \times 10^8 M_\odot$\\
Inner radius of the accretion-disk, $R_{\mathrm{AD}}^{\mathrm{in}}$ & $6 R_{\mathrm{G}}$\\
Outer radius of the accretion-disk, $R_{\mathrm{AD}}^{\mathrm{out}}$ & $10^3 R_{\mathrm{G}}$\\
Accretion disk luminosity, $L_{\mathrm{AD}}$ & $1 \times 10^{45}~ \mathrm{erg \cdot s^{-1}}$\\
Height of the emission region above & \\
the central black hole, $h$ & $1.4\times 10^{17}~\mathrm{cm}$\\
\enddata
\tablecomments{The parameters correspond to that of \cite{Paliya_etal2015} for 3C 279 in low-state. The seed photons are assumed to come directly from the disk, and the electron energy distribution is drawn from the electron fit spectrum of \cite{Baring_etal2017} in order to simulated the IC scattering off thermal shock-heated electrons in the jet of the blazar. \\
In the table above, $M_\odot \sim 1.9 \times 10^{33} \mathrm{g}$ is the solar mass, and $R_{\mathrm{G}} = GM_{\mathrm{BH}}/c^2$ is the gravitational radius of the black hole (with $G\sim 6.7 \times 10^{-8} \mathrm{cm^3 \cdot g^{-1} \cdot s^{-2}}$ the gravitational constant and $c \sim 3 \times 10^{10} \mathrm{cm\cdot s^{-1}}$ the speed of light).}
\end{deluxetable}

\subsection{The seed photon fields}\label{sec:SeedPhotons}
The primary seed photon fields for IC models of AGNs can either be external emission from the BLR and/or dusty torus (see e.g. \cite{Sikora_etal1994, Blazejowski_etal2000, Ghisellini_Tavecchio_2008}), or direct accretion-disk emission (see e.g. \cite{Dermer_Schlickeiser_1993, Bottcher_etal1997, Dermer_Schlickeiser_2002}). The MAPPIES code, therefore, draws the seed photons from either an isotropic, single-temperature black body distribution (corresponding to external emission from the BLR and/or dusty torus), or from a multi-temperature accretion-disk spectrum. In the first case, the radiation energy density $U_{\mathrm{DT}}^{\mathrm{AGN}}$ appears uniform in the AGN rest frame, provided that the emission region is located inside the characteristic scale of the BLR/dusty torus. The angular distribution in the AGN rest frame varies over angular scales $\Delta \theta \gg \Gamma_{\mathrm{jet}}^{-1}$ so that the angular distribution in the co-moving frame is dominated by relativistic aberration rather than intrinsic anisotropy \citep{Bottcher_etal2013}. The radiation energy density $U_{\mathrm{DT}}$ is thus boosted to a highly anisotropic field in the emission frame with $U_{\mathrm{DT}}\sim \Gamma_{\mathrm{jet}}^2U^{\mathrm{AGN}}_{\mathrm{DT}}$.

\begin{figure}[ht!]
\includegraphics[width=\columnwidth]{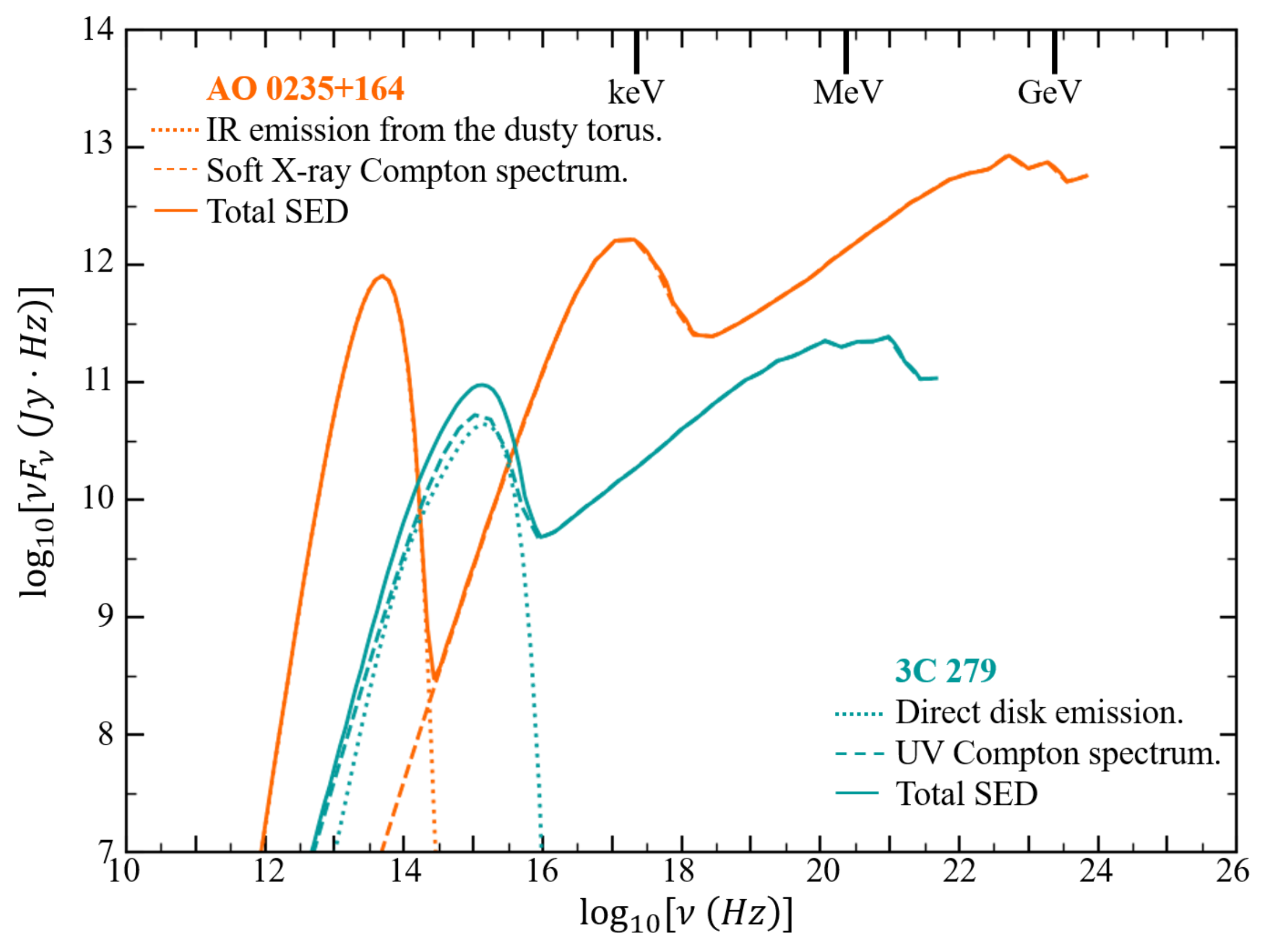}
\caption{The total SEDs for AO 0235+164 (solid orange curve) and 3C 279 (solid blue curve). The seed photon spectra are determined by the model parameters, and are assumed to be IR emission from the dusty torus for AO 0235+164 (dotted orange curve) and direct disk emission for 3C 279 (dotted orange curve). Compton scattering off thermal electrons in the jet (as shown in Figure \ref{fig:elecDist}) results in soft X-ray radiation for AO 0235+164 (dashed orange curve) and UV radiation for 3C 279 (dashed blue curve). \label{fig:ComptonSpectra}}
\end{figure}
Figure \ref{fig:ComptonSpectra} shows the seed photon spectra of AO 0235+164 (orange dotted curve) and 3C 279 (blue dotted curve). The seed photon field for AO 0235+164 is assumed to be IR emission from the dusty torus ($kT_{\mathrm{rad}} \sim 0.1 ~\mathrm{eV}$; dotted orange curve), analogous to the model used by \cite{Baring_etal2017}. The directly observable flux from the dusty is calculated from its luminosity, $L_{\mathrm{DT}} = (4 \pi R_{\mathrm{eff}}^2 c)\times U_{\mathrm{DT}}$, where $R_{\mathrm{eff}}$ is an effective size. The dusty torus is expected to have a characteristic radius of $R_{\mathrm{DT}} \sim$ a few pc. However, due to the torus-like geometry, its effectively IR emitting surface (as seen from a distant observer) is significantly smaller than $4 \pi R_{\mathrm{DT}}^2$. We therefore introduce an effective size of the dusty torus, $R_{\mathrm{eff}} \sim 6\times10^{17}~\mathrm{cm}$, which is expected to be about an order of magnitude smaller than $R_{\mathrm{DT}}$. The contribution of direct accretion-disk emission to the radiation density in the emission-region rest frame, is obtained as $U_{\mathrm{AD}}\sim 2.5 \times 10^{-5} ~\mathrm{erg \cdot cm^{-3}}$ from Equation 9 of \cite{Ghisellini_Madau_1996} for the combination of free parameters considered by \cite{Baring_etal2017}, and is negligible compared to that of isotropic IR emission from the dusty torus, where $U_{\mathrm{DT}} \sim \Gamma_{\mathrm{jet}}^2U_{\mathrm{DT}}^{\mathrm{AGN}} = 0.6 ~\mathrm{erg \cdot cm^{-3}}$ (with $U_{\mathrm{DT}}^{\mathrm{AGN}} \sim 5 \times 10^{-4} ~\mathrm{erg \cdot cm^{-3}}$ and $\Gamma_{\mathrm{jet}} = 35$). The emission region is assumed to be much closer to a more luminous accretion-disk for 3C 279 than in the case of AO 0235+164. The seed photons for 3C 279 are thus drawn from a multi-temperature accretion-disk spectrum (dotted blue curve), with a disk luminosity of $L_{\mathrm{AD}} = 1.0 \times 10^{45} ~\mathrm{erg \cdot s^{-1}}$ and a disk radius $R_{\mathrm{AD}} = R_{\mathrm{AD}}^{\mathrm{out}} - R_{\mathrm{AD}}^{\mathrm{in}} \sim 4.5 \times 10^{16} ~\mathrm{cm}$ (where $R_{\mathrm{AD}}^{\mathrm{out}}$ and $R_{\mathrm{AD}}^{\mathrm{in}}$ are the outer and inner radius of the disk, respectively).

\subsection{The electron energy distribution}\label{sec:Electrons}
Generally, the effects of Compton scattering depend on the electron energy distributions. DSA at relativistic shocks is thought to be an important acceleration mechanism in blazar jets, which may produce the non-thermal particles that emit the broad-band continuum detected from the jets. \cite{Baring_etal2017} employed the results from Monte-Carlo simulations of DSA at relativistic shocks by \cite{Summerlin_Baring_2012}, in order to model the particle acceleration at blazar shocks. The simulations captured the connection between the thermal component and the power-law tail of the non-thermal electrons in the blazar jet. The soft X-ray excess of AO 0235+164, modelled as IC scattering off the thermal, non-relativistic shock heated electrons, tightly constrained the energy dependence of the diffusion coefficients for the electrons. For more details on the model parameters and the electron energy distribution resulting in the SED fit, see \cite{Baring_etal2017}. 

\begin{figure}[ht!]
\includegraphics[width=\columnwidth]{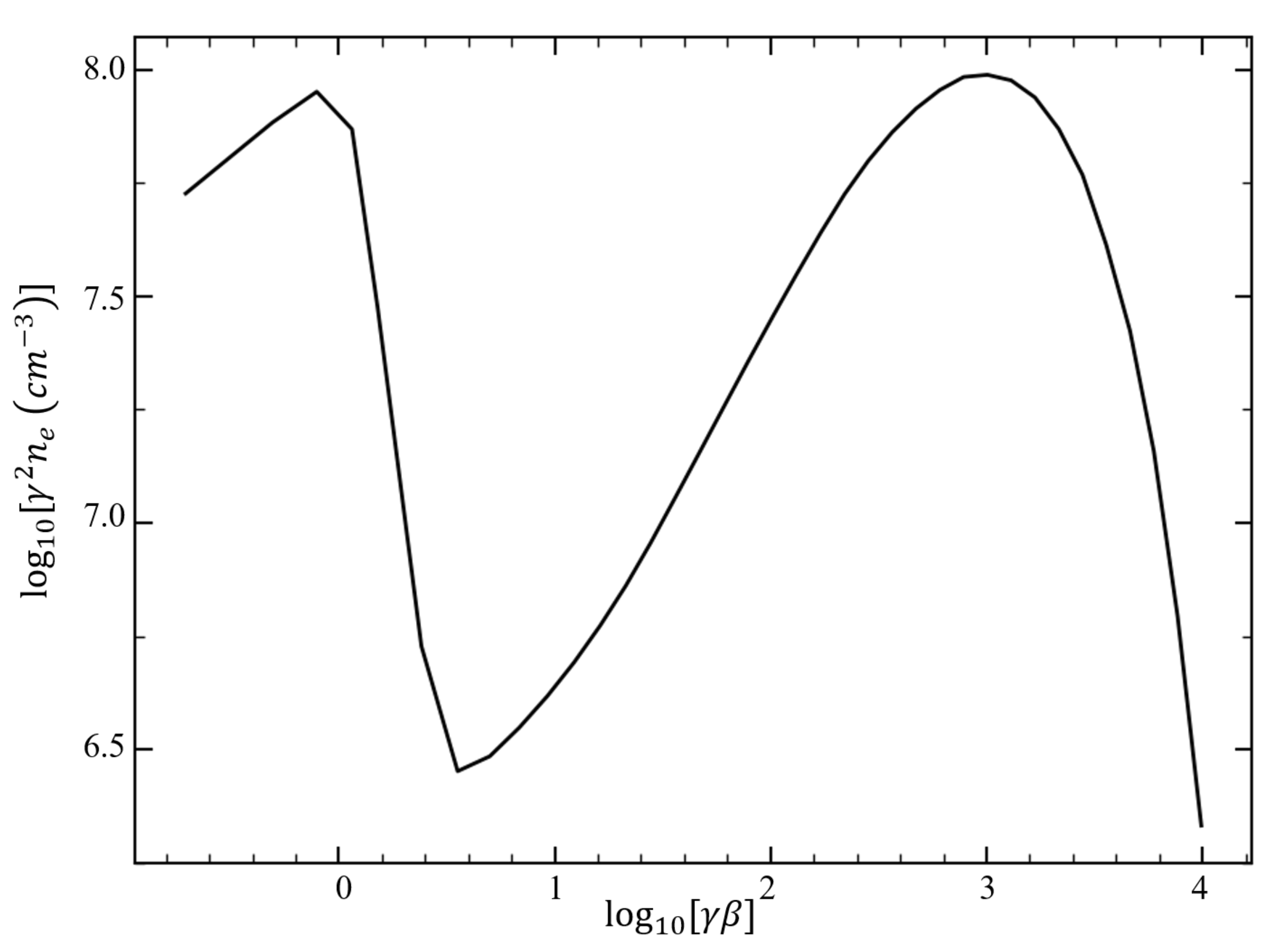}
\caption{The thermal and non-thermal electron distribution (with Lorentz factors of $\gamma$; drawn from the electron fit spectrum of \cite{Baring_etal2017}) as a function of the dimensionless electron momentum $\gamma\beta$. \label{fig:elecDist}}
\end{figure}
In the interpretation of the soft X-ray excess as a bulk Compton signature, a measurable degree of polarization in the frequency range in which the bulk Compton radiation dominates is expected, as it results from anisotropic Compton scattering off thermal, non-relativistic electrons. The MAPPIES code allows the quantification of this prediction and thereby, in comparison with future X-ray polarimetry data, the quantification of the presence of a pool of thermal electrons in the emission region of AO 0235+164. The electron energy distribution is thus drawn from the electron spectrum used in the SED fit by \cite{Baring_etal2017}, shown in Figure \ref{fig:elecDist}. The thermal-to-nonthermal particle ratio (comparable to that in the case of AO 0235+164) may result from shock acceleration of electrons in 3C 279 as well. Therefore, as an illustrative test case, we use the same electron distribution for 3C 279 as AO 0235+164 in order to study the potential bulk Comptonization effects.

\subsection{Polarized-dependent Compton scattering}
A variety of physical phenomena may alter the polarization state of the observed radiation, including the influence of the magnetic fields, general relativity, and the emission mechanisms. Synchrotron radiation of relativistic charged particles in ordered magnetic fields is expected to be both linearly and circularly polarized \citep{Westfold_1959, RybickiandLightman_1979}, while Compton scattering off relativistic electrons will reduce the degree of polarization to about half of the seed photon field's polarization \citep{Bonometto_etal1970}. For both blazar case studies, the seed photons are assumed to be unpolarized. Due to the polarization dependence of the Klein-Nishina cross section, polarized photons scatter preferentially in a direction perpendicular to their electrical field vector, and the electric field vectors of the scattered photons tend to align with the seed photons' electric field \citep{Matt_etal1996}. Polarization can therefore be induced when non-relativistic electrons scatter an anisotropic photon field, even if the seed photons are unpolarized.

The polarization signatures of the scattered photons are obtained by summing the photons' contribution to the Stokes parameters in a specified direction after the simulation is complete. Circular polarization may be generated either as an intrinsic component of Synchrotron radiation (i.e. if the seed photons are circular polarized Synchrotron emission) or via Faraday conversion of linear polarization into circular polarization driven by some internal Faraday rotation (see e.g. \cite{Wardle_etal1998, Homan_etal2009, MacDonald_Marscher_2018, Boehm_etal2019}). Since the seed photon fields are assumed to be unpolarized external radiation, and high-energy polarization
will not be affected by Faraday rotation due to the $\lambda^2$ (where $\lambda$ is the wavelength) dependence of this effect, we only consider the results of linear polarization. The results of circular polarization are unessential for making predictions of the expected polarization signatures for upcoming polarimetry missions, since all existing or proposed high-energy polarimeters only measure linear polarization. Due to the photon-counting nature of the X-ray and $\gamma$-ray observatories (and thus, polarimeters), to our knowledge, there is fundamentally no way to measure circular polarization in X-rays or $\gamma$-rays.

\section{Results and discussion}\label{sec:results}
In this section, the polarization signatures of IC scattering for the two blazar case studies will be discussed. The results of Compton scattering off a thermal population of shock-heated electrons contained in the jet of the blazar (as shown in Figure \ref{fig:elecDist}), are given for AO 0235+164 and 3C 279 for the combination of free parameters listed in Table \ref{tab:param_AO235_164} and Table \ref{tab:param_3C279}, respectively.
\subsection{The blazar SEDs}\label{sec:Compton}
Figure \ref{fig:ComptonSpectra} shows the SEDs of AO 0235+164 (orange) and 3C 279 (blue). The seed photon fields are shown in dotted lines (determined by the model parameters; see Section \ref{sec:SeedPhotons}), the IC spectra are shown in dashed lines, and the total (seed + IC) SEDs are shown in solid lines. For AO 0235+164, the seed photon field is assumed to be IR emission from the dusty torus (dotted orange curve). The photon frequency increases by a factor of $\sim \gamma^2 \Gamma_{\mathrm{jet}}^2$ 
\clearpage 
\begin{deluxetable*}{llcc}
\tablenum{3}
\tablecaption{Predictions of the expected PDs for future proposed polarimetry missions in a model where the UV/soft X-ray excess in the SEDs of AO 0235+164 and 3C 279 is due to a bulk Compton feature. \label{tab:PD_forMissions}}
\tablewidth{0pt}
\tablehead{
\colhead{\textbf{Blazar Case Study}} & \colhead{\textbf{Polarimeter}} & \colhead{\textbf{Frequency range} [$\mathrm{log_{10}(\nu/Hz)}$]} & \colhead{\textbf{PD} [\%]}
}
\startdata
\hline
{} & POLLUX & $14.4 - 15.5$ & $12 - 30$\\
{} & LAMP & $16.8$ & $43$ \\
{\textbf{AO 0235+164}} & REDSoX & $16.7 - 17.3$ & $30 - 46$ \\
{(IR emission from} & XPP & $16.6 - 18.3$ & $\lesssim 46$ \\
{the dusty torus)} & eXTP & $17.0 - 18.3$ & $\lesssim 40$ \\
{} & IXPE & $17.7 - 18.3$ & $\lesssim 20$ \\
{} & POLIX & $18.0 - 18.3$ & $\lesssim 10$ \\
\hline
{\textbf{3C 279}} & {POLLUX} & $14.4 - 15.5$ & $18 - 23$ \\
{(direct disk emission)} & & & \\
\hline 
\enddata
\tablecomments{The PDs in the table above are estimated from the results shown in the top panel of Figure \ref{fig:PDSpectra}, where the results for AO 0235+164 are shown in orange, and the results for 3C 279 are shown in blue. The frequency range listed in the table refers to the range where polarization could be detectable for the corresponding polarimetry missions.}
\end{deluxetable*}
\noindent (where $\Gamma_{\mathrm{jet}} = 35$, and $\gamma \sim 1$) due to Compton scattering off thermal electrons, which results in soft X-ray radiation (dashed orange curve). The soft X-ray spectrum peaks at $\mathrm{log_{10}(\nu/ Hz)}\sim 17.3$, in agreement with the detection of the soft X-ray excess in the SED of AO 0235+164 by \cite{Ackermann_etal2012}, and the results from \cite{Baring_etal2017} (as shown in Figure \ref{fig:Baring_etal2017}). For 3C 279, essentially all the seed photons (direct disk emission; dotted blue curve) enter the emission region from behind, since $(R_{\mathrm{AD}}/h) \sim 0.012 \ll \Gamma_{\mathrm{jet}}^{-1}$, causing the photons to receive a negative Doppler boost into the emission region frame. Compton scattering of the direct disk emission results, therefore, in UV radiation (dashed blue curve) with frequencies $\mathrm{log_{10}(\nu/ Hz)}\sim 14.0 - 16.0$, peaking at $\mathrm{log_{10}(\nu/ Hz)}\sim 15$, which is consistent with the detection of the UV excess by \cite{Pian_etal1999} and \cite{Paliya_etal2015}, as well as the reconstructed BBB of \cite{Roustazadeh_Botther_2012}.
\subsection{Compton polarization}\label{ComptonPol}
\begin{figure}[ht!]
\includegraphics[width=\columnwidth]{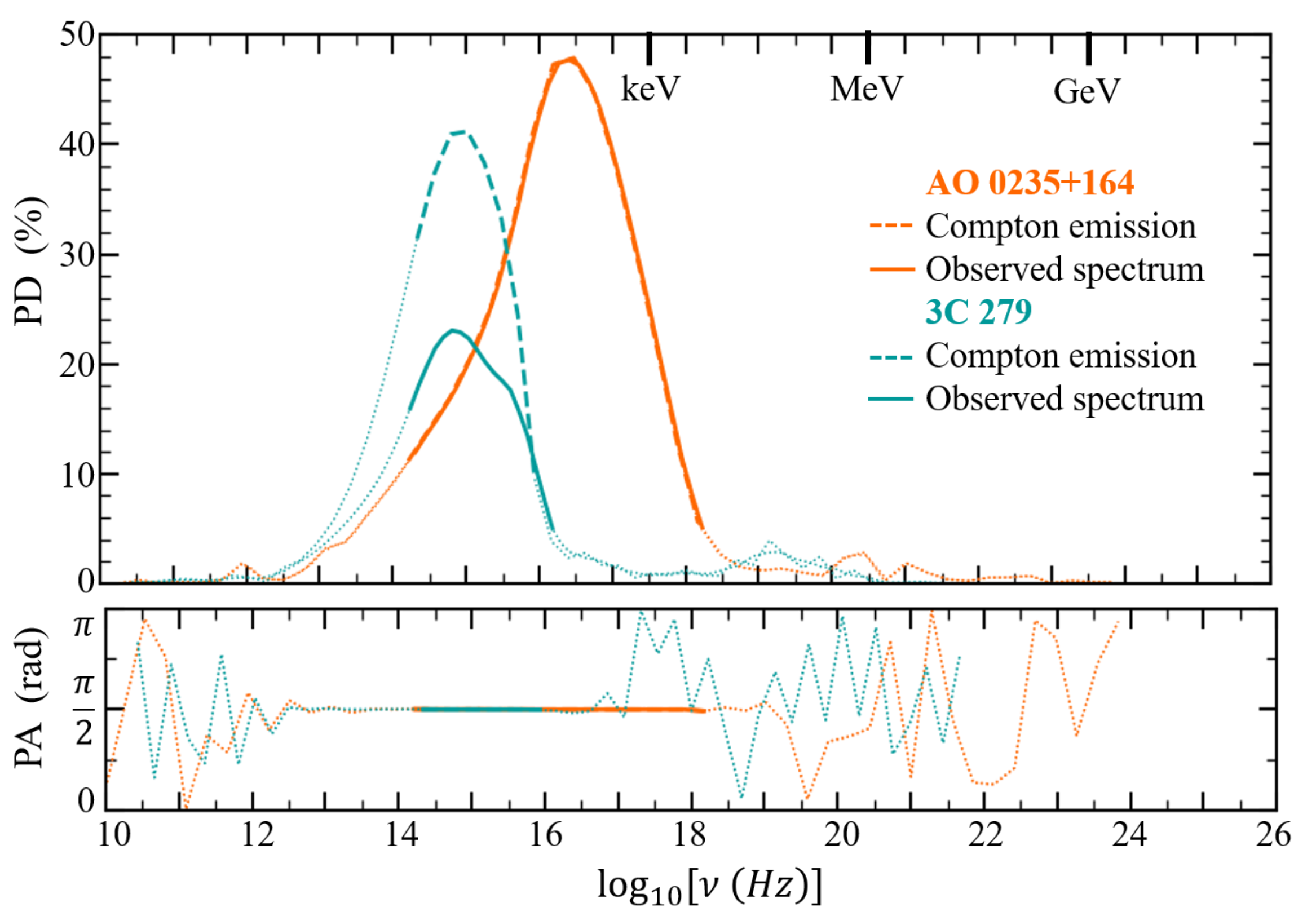}
\caption{The PD (top panel) and PA (bottom panel) as a function of the photon frequency for AO 0235+164 (orange) and 3C 279 (blue). The dotted lines indicate negligible polarization, and the frequency range in which other (electron-synchrotron) radiation components may dominate over the UV/soft X-ray components. The dashed lines indicate the polarization of the Compton emission, and the polarization of the total observed spectrum is shown with solid lines. \label{fig:PDSpectra}}
\end{figure}

The Compton polarization signatures are shown as a function of the photon frequency (in the observer's frame) for viewing angles of $\Theta_{\mathrm{AGN}} \sim \Gamma_{\mathrm{jet}}^{-1}~\mathrm{rad}$ in Figure \ref{fig:PDSpectra}. The Compton emission from AO 0235+164 (shown in orange) exhibits a significant PD (top panel) within the frequency range $\mathrm{log_{10}(\nu/ Hz)}\sim 14.3 - 18.3$ (restricted to the frequency range where the soft X-ray excess may dominate other -- electron synchrotron -- radiation components), with the maximum PD $\sim 48 \%$ at a frequency of $\mathrm{log_{10}(\nu/ Hz)}\sim 16.4$, and a PD of $\sim 30 \%$ at the peak of the soft X-ray component (at $\mathrm{log_{10}(\nu/ Hz)}\sim 17.3$). The frequency range in which significant polarization is predicted, covers a number of upcoming and proposed missions for UV and X-ray polarimetry, which include POLLUX on board the Large UV/Optical/IR Surveyor (LUVOIR; \cite{POLLUX_2019}), the Lightweight Asymmetry and Magnetism Probe (LAMP; \cite{LAMP_2019}), the Rocket Experiment Demonstration of a Soft X-ray Polarimeter (REDSoX; \cite{REDSoX_2019}), the $X$-ray Polarization Probe (XPP; \cite{XPP_2019b}), the Enhanced $X$-ray Timing and Polarimetry Mission (eXTP; \cite{eXTP_2019}), the Imaging $X$-ray Polarimetry Explorer (IXPE; \cite{IXPE_2020}), and the $X$-ray Polarimeter Experiment (POLIX; \cite{POLIX_2016}). The predicted PDs are listed in Table \ref{tab:PD_forMissions}, and are expected to decrease with increasing frequency for all the polarimeters considered except POLLUX, for which the PD is expected to increase with increasing frequency. 

For 3C 279, the Compton polarization (shown in blue) is detectable in the frequency range of $\mathrm{log_{10}(\nu/ Hz)}\sim 14.3 - 16.0$ (restricted to the frequency range where the BBB may make a significant contribution to other -- electron synchrotron and direct accretion-disk -- radiation components).  The total observed spectrum has PDs up to $23 \%$ (solid blue curve), which is about half of the bulk Compton emission polarization (dashed blue curve) due to the contribution of the unpolarized direct disk emission. The UV polarimeter POLLUX could thus be able to detect the polarization of the BBB in the SED of 3C 279 within the frequency range $\mathrm{log_{10}(\nu/ Hz)}\sim 14.4 - 15.5$, with expected PDs of $(18 - 23)\%$ (see Table \ref{tab:PD_forMissions}), with the maximum PD $\sim 23\%$ at the peak of the BBB (at $\mathrm{log_{10}(\nu/ Hz)}\sim 15$). 

\begin{figure}[ht!]
\includegraphics[width=\columnwidth]{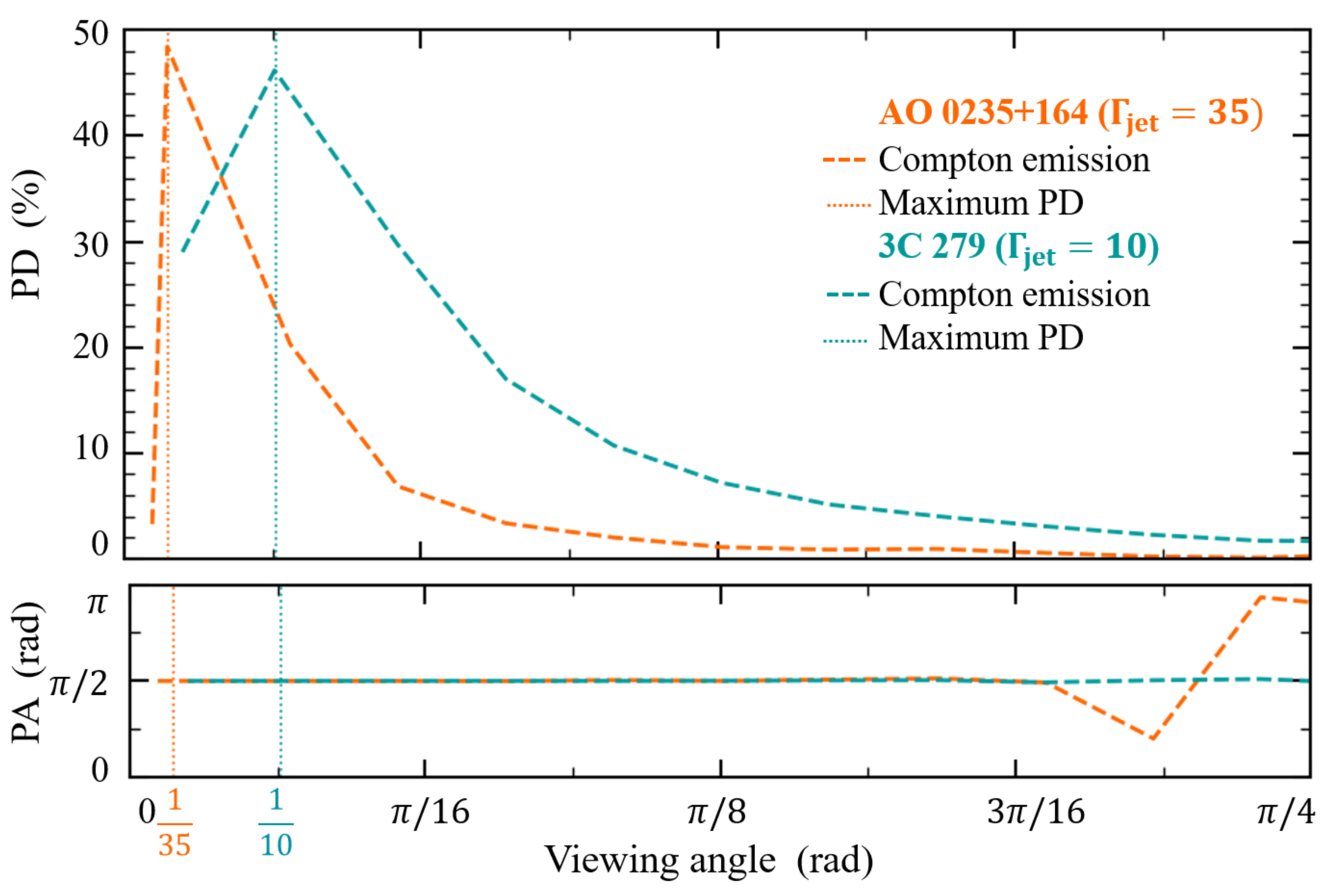}
\caption{The PD (top panel) and PA (bottom panel) as a function of the scattered photon viewing angle (in the AGN-rest frame) for AO 0235+164 (orange) and 3C 279 (blue). The dotted lines indicate the viewing angle where the maximum polarization occurs. \label{fig:PDDist}}
\end{figure}
The Compton polarization signatures are shown as a function of the scattered photon viewing angle $\Theta_{\mathrm{AGN}}$ in Figure \ref{fig:PDDist}. In the emission region rest frame, most of the seed photons move in the negative jet direction. The maximum polarization, therefore, occurs at the right angle, $\Theta_{\mathrm{AGN}} \sim (\pi/2) ~\mathrm{rad}$, in the electron rest frame, which is essentially the same in the emission region rest frame. Boosting to the AGN rest frame, the maximum PD occurs at $\Theta_{\mathrm{AGN}} \sim \Gamma_{\mathrm{jet}}^{-1} ~\mathrm{rad}$, with $\Gamma_{\mathrm{jet}} = 35$ for AO 0235+164 (indicated with a orange dotted line) and $\Gamma_{\mathrm{jet}} = 10$ for 3C 279 (indicated with a blue dotted line). The polarization angle (PA; shown in the bottom panels of Figures \ref{fig:PDSpectra} and \ref{fig:PDDist}) of the polarized fraction of the Compton emission assumes a constant value of PA $= (\pi/2) ~\mathrm{rad}$ for both blazar case studies, which refers to polarization perpendicular to the jet axis.

\section{Summary and conclusion}\label{sec:summary}
In this paper, the MAPPIES code is used to simulate IC scattering off a thermal population of shock-heated electrons contained in the blazar jet. Predictions of the polarization signatures, in a model where the UV/soft X-ray excess in blazar spectra is due to bulk Comptonization of external radiation fields, as proposed by \cite{Baring_etal2017}, are made for AO 0235+164 and 3C 279. Compton scattering of external IR emission from the dusty torus results in soft X-ray radiation for AO 0235+164, and Compton scattering of UV emission from the accretion-disk results in UV radiation for 3C 279 (as shown in Figure \ref{fig:ComptonSpectra}). The Compton X-ray spectrum of AO 0235+164 agrees with the results of \cite{Baring_etal2017}, and the UV spectrum of 3C 279 is consistent with the BBB detected by e.g. \cite{Pian_etal1999, Paliya_etal2015} in the low state. Therefore, while an isotropic IR radiation field (in the AGN rest frame) is required to reproduce the soft X-ray excess as bulk Compton emission in the SED of AO 0235+164, direct disk emission likely dominates the seed radiation field for 3C 279, with the emission region closer to a more luminous disk compared to that of AO 0235+164. 

The thermal Comptonization process involved in the bulk Compton feature leads to significant polarization within the UV/soft X-ray excess in the SEDs of both blazar case studies (PD $\lesssim 48\%$ for AO 0235+164 and PD $\lesssim 23\%$ for 3C 279; as shown in Figure \ref{fig:PDSpectra}). The maximum PD occurs at viewing angles of $\Theta_{\mathrm{AGN}} \sim \Gamma_{\mathrm{jet}}^{-1} ~\mathrm{rad}$ (shown in Figure \ref{fig:PDDist}), and the PA for the polarized fraction of the Compton emission assumes a constant value of PA$=(\pi/2) ~\mathrm{rad}$, which corresponds to polarization perpendicular to the jet axis. The Compton polarization of the emission from the UV/soft X-ray excess in the SEDs is predicted to be detectable within the frequency range of $\mathrm{log_{10}(\nu/ Hz)}\sim 14.3 - 18.3$ for AO 0235+164, and $\mathrm{log_{10}(\nu/ Hz)}\sim 14.3-16.0$ for 3C 279. Future missions anticipating to deliver polarization measurements of UV and X-ray emission from blazar jets may thus be able to detect the Compton polarization of the UV/soft X-ray excess in the SEDs of AO 0235+164 and 3C 279, with the expected PDs listed in Table \ref{tab:PD_forMissions}. This reinforces future prospects of using the polarization in the UV and X-ray regime, combined with spectral fitting and variability, in order to probe different models of the emission mechanisms responsible for different spectral features in blazar spectra.
\section*{Acknowledgements}
The work of Markus B{\"o}ttcher is supported through the South African Research Chair Initiative of the National Research Foundation \footnote{Any opinion, finding and conclusion or recommendation expressed in this material is that of the authors and the NRF does not accept any liability in this regard.} and the Department of Science and Innovation of South Africa, under SARChI Chair grant No. 64789.
\bibliography{bib}{}
\bibliographystyle{aasjournal}
\end{document}